\documentclass[12pt,preprint]{aastex}

\slugcomment{This manuscript is accepted by \textbf{Ap\&SS}}

\shorttitle{isothermal IRDCs}

\shortauthors{Bahmani and Nejad-Asghar}

\begin{document}

\title{The effect of magnetic field morphology on the structure of massive IRDC clumps}

\author{Nahid Bahmani, Mohsen Nejad-Asghar}

\affil{Department of Physics, University of Mazandaran, Babolsar,
Iran}

\email{nejadasghar@umz.ac.ir}

\begin{abstract}
Infrared dark clouds (IRDCs) have dense elongated clumps and
filaments with the favorable viewing condition of being on the
near-side of a bright mid-infrared background. The clumps usually
have multiple cores \textit{around the center}. In this work, we
study the effect of magnetic field morphology on the structure of
massive IRDC clumps. To achieve this goal, we consider an
axisymmetric isothermal oblate IRDC clump, embedded into a constant
external magnetic field. We assume a polynomial function for the
magnetic field morphology inside the clump. We use the numerical
iterative methods to solve the equations: the successive
over-relaxation method to find the magnetic and gravitational
fluxes, and then the bicongugate gradient method to find the
optimized values of mass and current densities. The results show
that the IRDC clump will be very elongated along the perpendicular
direction of the external magnetic field lines. Also, the assumption
of choosing of a polynomial function for the magnetic field
morphology leads to the formation of dense regions \textit{around
the center}. The greater the density of the central region, the
larger the density of these dense regions and the closer to the
center. The presence of these dense regions can lead to the
formation of cores at these points.
\end{abstract}


\section{Introduction}

Infrared dark clouds (IRDCs) are cold, dense molecular clouds seen
silhouetted against the bright diffuse mid-infrared emission of the
Galactic plane. They were discovered during mid-infrared imaging
surveys with the Infrared Space Observatory (P\'{e}rault et
al.~1996) and Midcourse Space Experiment (Egan et al.~1998). Simon
et al.~(2006) used 8.3 $\mu m$ mid-infrared images acquired with the
Midcourse Space Experiment satellite to identify and catalog IRDCs
in the first and fourth quadrants of the Galactic plane. The
observations show that the IRDCs have low temperatures $(<25\,
\mathrm{K})$, high densities $(>10^{5}\, \mathrm{cm^{-3}})$ and
sizes of $1-10\, \mathrm{pc}$  (e.g., Carey et al.~1998, 2000,
Butler \& Tan~2009,Chira et al.~2013, Feng et al.~2016).

Many of these IRDCs have the filamentary structure with high mass
non-spherical clumps on $0.01-1\, \mathrm{pc}$ size scales; two
examples are G028.37+00.07 (e.g. Lim \& Tan~2014) and G035.39-00.33
(e.g., Sokolov et al.~2017). The masses of these non-spherical IRDC
clumps are generally of order hundreds to thousands of solar masses,
and they are believed to be the precursors for formation of the
massive stars and/or stellar clusters. The IRDC clumps exhibit a
variety of star and cluster formation stages: from prestellar, dark,
cold, and quiescent cores to active, infrared-bright and chemically
rich substructures with embedded sources driving outflows and HII
regions (e.g., Pillai et al.~2006, Chambers et al.~2009, Battersby
et al.~2010, Sanhueza et al.~2012, Wang et al.~2011, 2014).

Most of the IRDC clumps have elongated shapes (e.g., recent report
of Sanhueza et al.~2017 in the IRDC G028.23-00.19). As if a static
gaseous cloud is only affected by its self-gravitational force, the
spherical symmetry requires that it be a sphere. The observed
non-spherical shapes like as IRDC clumps are created either by the
non-static effects or by the non-symmetric forces such as the
magnetic forces. The IRDC clumps are assumed to be near virial
equilibrium and in approximate pressure equilibrium with the
surrounding environment. A massive, virialized non-spherical IRDC
clump in pressure equilibrium with its environment cannot be
supported by thermal pressure, given observed temperatures of
$T<25\, \mathrm{K}$, and so must be supported by some form of
non-thermal pressure, e.g., magnetic fields. The magnetic field
pressure can lead to form some elongated shapes that are
perpendicular to the magnetic field lines (e.g., Mouschovias~1976).
Estimates of the magnetic field strengths in the IRDCs indicate
values in the range $100-700\, \mu G$ (e.g., Santos et al.~2016,
Henshaw et al.~2016, Hoq et al.~2017).

The observations show that the IRDC clumps have many prestellar and
protostellar cores. For example, Sanhueza et al.~(2017) have
recently reported some observational information about a massive
elongated clump in the IRDC G028.23-00.19. This clump hosts many
cores around its elongated axis (like SMA1 and SMA4 \textit{around
the center}). Hosting the cores \textit{around the center} and along
the elongated axis of the IRDC clumps may lead to the formation of
non-spherical stellar clusters. Characterizing the initial
conditions for formation of the stellar clusters is important for
distinguishing between various theoretical models.

Formation of the multiple cores along the major axis of a massive
IRDC clump may be justified via some theoretical works that
describing the fragmentation of hydrodynamic fluid cylinders (e,g,.
Chandrasekhar and Fermi~1953, Tomisaka~1995, Hernandez and Tan~2011,
Contreras et al.~2016). Henshaw et al.~(2016) investigated the
spatial distribution of the cores along the major axis of the IRDC
G035.39-00.33. They found a significant discrepancy between the
observed distribution of the cores, and that predicted by the
theoretical works, which used the fragmentation of hydrodynamic
fluid cylinders. Also, these theoretical models cannot justify the
distribution of the cores \textit{around the center} of an IRDC
clump. Can the presence of a magnetic field through the IRDC clump,
in addition to its force for formation of non-spherical shape, also
justify some constraints on the spatial positions of the cores in
them? Here, we want to present a theoretical model for imprint of
the morphology of the magnetic field lines to justify formation of
the cores through elongated axis (and \textit{around the center}) of
a massive IRDC clump. For this purpose, the formulation of the
theoretical model is given in \S~2. The boundary conditions and the
morphology of the magnetic field is given in \S3. The results are
presented at \S4, and section~5 is devoted to a summary and
conclusions.

\section{Formulation of the problem}

We consider a nearly spherical (oblate) massive IRDC clump, which is
in the isothermal state with sound speed $c_{s}$. We use axisymetric
cylinder polar coordinate $(\varpi,z)$, centered on the origin of
this IRDC clump. The magnetic field is assumed to be uniform in the
outer regions of the clump. The $z$ direction is considered along
this uniform magnetic field. We must consider a suitable morphology
of the magnetic field in the inner regions of the IRDC clump. The
schematic diagram of the shape of the IRDC clump and its magnetic
field configuration is shown in the Fig.~\ref{magnetic}.

The $\Phi_{B}$ is the magnetic flux contained within the surface
generated by rotating any magnetic field line about the $z$-axis
(Fig.~\ref{magnetic}). The relation $\nabla.\mathbf{B}=0$ allows us
to recast $\mathbf{B}$ in terms of the magnetic potential
$\mathbf{A}$ (i.e., $\mathbf{B}=\nabla\times \mathbf{A}$). The
vector field $\mathbf{A}$ need have only a nonzero $\phi$-component
in order to generate an arbitrary poloidal $\mathbf{B}$-field. If
$S$ be any circular two-dimensional surface centered on the $z$-axis
and bounded by a circle of radius $\varpi$, then $\Phi_{B}=\int_S
(\nabla\times \mathbf{A}). \hat{k}\, da = 2\pi\varpi A_{\phi}$.
Employing the unit vector ${\hat{e}_{\phi}}$, we write
\begin{eqnarray}\label{magn1}
  \nonumber \textbf{B} &=& \nabla\times(A_{\phi}{\hat{e}_{\phi}}) \\
   \nonumber &=& -\frac{{\hat{e}_{\phi}}}{\varpi}\times\nabla(\varpi A_{\phi})\\
  &=&-\frac{{\hat{e}_{\phi}}}{2\pi\varpi c}\times\nabla\Phi_{B}.
\end{eqnarray}
By using of Ampere's law $\nabla\times \mathbf{B}=\frac{4 \pi}{c}
\mathbf{j}$ and equation~(\ref{magn1}), we obtain the magnetic
induction equation as
\begin{equation}\label{magn}
 \frac{\partial}{\partial\varpi}\left(\frac{1}{\varpi}\frac{\partial\Phi_{B}}{\partial\varpi}\right)+\frac{1}{\varpi}\frac{\partial^2\Phi_{B}}{\partial
 z^2}=-\frac{8\pi^2}{c}j_{\phi},
\end{equation}
where $j_{\phi}$ is the $\phi$-component of current density.

The equation for force balance is
\begin{equation}\label{force}
    0=-c^{2}_{s}\nabla\rho-\rho\nabla\Phi_{g}+\frac{1}{c} \mathbf{j} \times
    \mathbf{B},
\end{equation}
where $\Phi_g$ is the gravitational potential
\begin{equation}\label{poas}
 \frac{1}{\varpi}\frac{\partial
 }{\partial\varpi}\left(\varpi\frac{\partial\Phi_{g}}{\partial\varpi}\right)+\frac{\partial^2\Phi_{g}}{\partial
 z^2}=4\pi G\rho.
\end{equation}
With poloidal magnetic field and toroidal current density, the
magnetic force per unit volume, $\frac{1}{c} \mathbf{j}\times
\mathbf{B}$, must be a poloidal vector with both $\varpi$ and $z$
components. In the this way, the force balance
equation~(\ref{force}) can be rewritten as
\begin{equation}\label{force1}
    c^{2}_{s}\frac{\partial \rho}{\partial \varpi}+ \rho \frac{\partial \Phi_{g}}{\partial \varpi}=\frac{j_{\phi}}{2\pi\varpi c}\frac{\partial\Phi_{B}}{\partial\varpi},
\end{equation}
\begin{equation}\label{force2}
   c^{2}_{s} \frac{\partial \rho}{\partial z}+\rho\frac{\partial\Phi_{g}}{\partial z}=\frac{j_{\phi}}{2\pi\varpi c}\frac{\partial \Phi_{B}}{\partial
   z}.
\end{equation}

We use the scale values of length, temperature, and density equal to
$[l]=1\,\mathrm{pc}$, $[T]=10\, \mathrm{K}$, $[\rho]=2 m_H \times
(10^4\, \mathrm{cm^{-3}})$, respectively, so that the scale of
magnetic field strength is $[B]=18.6\, \mathrm{\mu G}$ and the mass
scale is $[M]=9850\, \mathrm{M_{\odot}}$. Using these scale units,
the basic equations for the axisymmetric massive isothermal IRDC
clump in the force balance equilibrium can be rewritten as
\begin{equation}\label{neweq4}
    \frac{\partial}{\partial\varpi}\left(\frac{1}{\varpi}\frac{\partial\Phi_{B}}{\partial\varpi}\right)+\frac{1}{\varpi}\frac{\partial^2\Phi_{B}}{\partial
 z^2} = -j_{\phi},
\end{equation}
\begin{equation}\label{neweq3}
    \frac{1}{\varpi}\frac{\partial
 }{\partial\varpi}\left(\varpi\frac{\partial\Phi_{g}}{\partial\varpi}\right)+\frac{\partial^2\Phi_{g}}{\partial
 z^2} = \rho,
\end{equation}
\begin{equation}\label{neweq1}
   \frac{\partial\rho}{\partial\varpi}+\rho\frac{\partial\Phi_{g}}{\partial\varpi} = \frac{j_{\phi}}{\varpi}
   \frac{\partial\Phi_{B}}{\partial\varpi},
\end{equation}
\begin{equation}\label{neweq2}
    \frac{\partial\rho}{\partial z}+\rho\frac{\partial\Phi_{g}}{\partial z} = \frac{j_{\phi}}{\varpi} \frac{\partial\Phi_{B}}{\partial
    z}.
\end{equation}
For solving these equations and determining the internal profile
structure of an IRDC clump, we need to know the boundary values
(i.e., values of $\rho$, $j_\phi$, $\Phi_B$, and $\Phi_g$ on the
dash lines~$1-4$ of Fig.~\ref{magnetic}).

\section{Values in the boundaries}

We consider an isothermal IRDC clump with temperature $T\approx 1\,
[T]$ and total mass $M\approx 0.15\, [M]$. The clump is embedded in
the inter-clump medium of IRDC. The inter-clump medium is assumed to
have density $\rho_{int}\approx 1\, [\rho]$ and uniform magnetic
field $B_{int}\approx 8.5\, [B]$. These values are typically
correspond to the MM1 clump through the IRDC G28.23-00.19 (Sanhueza
et al.~2017, Hoq et al.~2017). The radii of the clump in the
$z$-axis and $\varpi$-axis directions are $R_0$ and $\lambda R_0$,
respectively, where $\lambda>1$ is the oblate axis ratio. For
simplicity, we assume that the boundaries at one quadrant of $\varpi
- z$ plane, are given as four lines~$1-4$, as depicted in the
Fig.~\ref{magnetic}.

In boundary line~$1$ with $(\varpi=0,z)$, the $j_\phi$ and $\Phi_B$
are zero so that the equation~(\ref{neweq2}) imply that $\rho =
\rho_c \exp(-\Phi_g)$, where $\rho_c$ is the central density in unit
of $[\rho]$. Here, we assume that the density on this boundary line
is corresponded with the profile of an infinite layer. In the
assumption of infinite layer, the gravitational potential is
independent of $\varpi$ so that the equation~(\ref{neweq3}) becomes
\begin{equation}\label{line1phig}
   \frac{d^2\Phi_g}{d z^2} = \rho_c \exp(-\Phi_g).
\end{equation}
This equation can be solved numerically with boundary conditions
$\Phi_{g(\varpi=0,z=0)} = 0$ and
$\frac{d\Phi_g}{dz}|_{(\varpi=0,z=0)} =0$. By choosing each value
for $\rho_c> \rho_{int}$, there is a special value for $R_0$ in
which the density at $z=R_0$ will be equal to $\rho_{int}$. The
maximum allowed radius of the IRDC clump in the $z$-axis direction,
$R_0$, for different values of the central density, $\rho_c$, is
shown in the Fig.~\ref{rhocR0}. The density contrast,
$\rho_c/\rho_{int}$, greater than $3.5$ causes to contract the layer
in the $z$-axis direction due to the importance of the
self-gravitational force. Counterpart of this case, in the
spherically symmetric clouds, is known as the occurrence of
gravitational instability (i.e., Bonnor-Ebert spheres, see, e.g.,
Stahler and Palla~2004, Fig.~9.2 for isothermal case,
Nejad-Asghar~2016, Fig.~3 for non-isothermal cases). Here, we choose
the boundary line~$4$ being near the maximum allowed value of $R_0$
(i.e., $Z_\infty \approx 0.94$). To consider the oblateness of the
IRDC clump and departure from infinite layer approximation, we will
approximately consider the term $\frac{1}{\varpi}\frac{\partial
}{\partial\varpi}\left(\varpi\frac{\partial
\Phi_{g}}{\partial\varpi}\right)$, and reevaluate the $R_0$, in each
iterative process as mentioned in the next section.

In boundary line~$2$ with $(\varpi,z=0)$, we assume the density and
the magnetic field strength are corresponded with the profiles of an
infinite cylinder. In the assumption of infinite cylinder, the
quantities $\Phi_g$, $\rho$, $j_\phi$, and $\Phi_B$ are independent
of $z$ so that the basic equations (\ref{neweq4})-(\ref{neweq2})
reduce to
\begin{equation}\label{line2}
  \frac{d^2\rho}{d\varpi^2}
  +\left(\frac{1}{\varpi}+\frac{f}{\rho}\right)\frac{d\rho}{d\varpi}-\frac{1}{\rho}\left(\frac{d\rho}{d\varpi}\right)^2+\rho^2
  =\frac{f}{\varpi}+\frac{df}{d\varpi},
\end{equation}
where $f \equiv 2\pi j_\phi B$. Knowing the magnetic field $B$ in
this boundary line, the magnetic flux and current density can be
obtained via $\Phi_B = \int_0^\varpi 2\pi \varpi' B d\varpi'$ and
$j_\phi = -2\pi \frac{dB}{d\varpi}$, respectively. If we choose a
suitable model for the magnetic field $B$, the
equation~(\ref{line2}) can be solved numerically with boundary
conditions $\rho_{(\varpi=0,z=0)} = \rho_c$ and $\frac{d \rho}{d
\varpi}|_{(\varpi=0, z=0)} = 0$. The oblate axis ratio $\lambda$ is
identified so that $\rho_{(\varpi= \lambda R_0, z=0)} = \rho_{int}$.
Then, the gravitational potential can be obtained from
equation~(\ref{neweq1}) as $\Phi_g = -\ln \frac{\rho}{\rho_c} +
\int_0^\varpi \frac{f}{\rho} d\varpi'$.

The most important thing in the boundary line~$2$ is the morphology
of the magnetic field. We do not have enough observational
information about the magnetic field strengths through the sub-pc
scales of the IRDC clumps (Hoq et al.~2017). We assume that the
magnetic field at the center of the IRDC clump is $(1+\eta) B_{int}$
where $\eta$ is the fractional change of the magnetic field strength
at the central region. If we use a power law relation(i.e., $B
\propto \rho^\kappa$) between the central magnetic field strength
and the central density (Crutcher~2012), we have $\eta = \left(
\frac{\rho_c}{\rho_{int}} \right)^\kappa -1$. Here, we choose
$\kappa\approx0.3$ (Nejad-Asghar~2016). Here, we consider a
polynomial function for the magnetic field morphology. The first
term that justify the condition $j_\phi = -2 \pi
\frac{dB}{d\varpi}|_{(\varpi=0,z=0)}=0$ is the quadratic term as
$B_{(\varpi,z=0)} =B_{int} \times \left[(1+\eta)-a_2
\varpi^2\right]$, where the coefficient $a_2 \geq 0$ determines the
slope of decreasing of the magnetic field by moving away from the
center. Since the slope is negative, the magnetic force is in the
positive $\varpi$ direction. This force implies that a pressure
gradient must be occurred so that the force balance (\ref{force}) be
justified. In the absence of magnetic field, the force balance
between the gravitational force and pressure gradient lead to
decreasing of pressure (and also decreasing of density in the
isothermal case) by moving away from the center. The magnetic force
is in the inverse direction of the gravitational force, thus,
considering of the magnetic field morphology with greater negative
slopes (i.e., grater values of $a_2$) can lead to increasing of the
pressure (and also density) by moving away from the center. Solution
of the equation (\ref{line2}), near the origin, for different values
of $a_2$ are shown in Fig.~\ref{rho-varpi}. The next boundary
conditions on the magnetic strength are $j_{\phi(\varpi= \lambda
R_0,z=0)}= 0$ and $\Phi_{B(\varpi= \lambda R_0,z=0)}= \pi (\lambda
R_0)^2 B_{int}$. Considering the next next term of the polynomial
function (i.e., cubic term) leads to choosing the magnetic
morphology as
\begin{equation}\label{Bl2}
B_{(\varpi,z=0)}= B_{int} \times
       \left[ (1+\eta)-\frac{10}{7} \left( 3\eta\frac{\varpi^2}{\lambda^2 R_0^2}
       - 2\eta \frac{\varpi^3}{\lambda^3 R_0^3} \right) \right].
\end{equation}
In the region $\varpi > \lambda R_0$, the magnetic field will
increase slowly to reach the inter-clump value $B_{int}$. This
region is not important for us.

In boundary lines~$3$ and $4$, which are assumed to be at the
inter-clump medium, we have $\rho= \rho_{int}$ and $j_\phi=0$. The
magnetic flux at the boundary line~$3$ is $\Phi_B = \pi (\lambda
R_0)^2 B_{int}$ and at the boundary line~$4$ is $\Phi_B = \pi
\varpi^2 B_{int}$. For gravitational potential, Mouschovias~(1976)
used a point mass approximation located at the origin of
coordinates. Since, oblateness of a spherical mass gives it a
nonzero quadrupole moments (e.g., Fitzpatrick~2012), here we add the
quadrupole approximation as
\begin{equation}\label{phig34}
  \Phi_{g(\varpi,z)}=J_0\frac{M}{(z^2+\varpi^{2})^{\frac{1}{2}}}
  +J_2\frac{M R_0^{2}}{(z^2+\varpi^2)^{\frac{3}{2}}},
\end{equation}
where $J_0$ and $J_2$ are chosen so that the continuity of the
gravitational potential being established at the intersection of
boundary lines~$1$ and $4$, i.e., $(\varpi=0,z=Z_\infty)$, and the
intersection of boundary lines~$2$ and $3$, i.e., $(\varpi=\lambda
R_0, z=0)$. If $\Phi_{g14}$ is the result of gravitational potential
at $(\varpi=0,z=Z_\infty)$ obtained from integration across the
boundary line~$1$, and $\Phi_{g23}$ is the result of gravitational
potential at $(\varpi=\lambda R_0, z=0)$ obtained from integration
across the boundary line~$2$, we have
\begin{eqnarray}\label{J0J2}
  \nonumber J_0= \frac{1}{\lambda^2-\alpha^2} \frac{(\lambda^3
  \Phi_{g23}-\alpha^3 \Phi_{g14})R_0}{M},\\
  \nonumber J_2= - \frac{\alpha^2\lambda^2}{\lambda^2-\alpha^2} \frac{(\lambda
  \Phi_{g23}-\alpha\Phi_{g14})R_0}{M},
\end{eqnarray}
where $\alpha \equiv Z_\infty/R_0$.

\section{Results}

In the previous section, the boundary values are specified around
the edge of the rectangular grid. In the interior points, the
equations (\ref{neweq4})-(\ref{neweq2}) must be satisfied. We
consider these equations by the finite-difference method, and employ
an iterative process to find the values of $\rho$, $\Phi_{g}$,
$j_{\phi}$ and $\Phi_{B}$ at the internal grid points.

First, we consider the equations (\ref{neweq4}) and (\ref{neweq3})
with assumption that the values of $\rho$ and $j_\phi$ are known at
the all (i.e., known fixed values at boundaries and assumed guessed
values at internal) grid points. The best initial guess for $\rho$
and $j_\phi$, at an internal point $(i,j)$, is a linear functional
form as
\begin{equation}\label{linerhoj}
   f_{i,j} = f_{i,1} + \frac{f_{i,N} - f_{i,1}}{Z_\infty} z_j,
\end{equation}
where $f$ corresponds to $\rho$ and $j_\phi$, and the integer
numbers $i$ (corresponding to the $\varpi$-axis) and $j$
(corresponding to the $z$-axis) run from $2$ to a maximum value
$N-1$. The finite difference of the equations (\ref{neweq4}) and
(\ref{neweq3}), on the rectangular grid, is an equation of the form
\begin{equation}\label{sor}
   a_{i,j} u_{i+1,j} +b_{i,j} u_{i-1,j} +c_{i,j} u_{i,j+1} +d_{i,j} u_{i,j-1} +e_{i,j} u_{i,j} = f_{i,j},
\end{equation}
where $u$ corresponds to $\Phi_B$ and $\Phi_g$, and the coefficients
$a$, $b$, $c$, $d$ and $e$ are all known values on the internal grid
points. We use the successive-over-relaxation (SOR) method (Press et
al.~2007) to solve the equation~(\ref{sor}), and to find the values
of $\Phi_B$ and $\Phi_g$ on the internal grid points.

Now, we turn our attention to the equations (\ref{neweq1}) and
(\ref{neweq2}) to find some best estimated new values for $\rho$ and
$j_\phi$. Eliminating $j_\phi$ between (\ref{neweq1}) and
(\ref{neweq2}) leads to a finite difference equation for density
$\rho$ as follows
\begin{equation}\label{newrho}
   \frac{f^{Bz}_{i,j}}{2 \Delta \varpi} \rho_{i+1,j} -
   \frac{f^{Bz}_{i,j}}{2 \Delta \varpi} \rho_{i-1,j} -
   \frac{f^{B\varpi}_{i,j}}{2 \Delta z} \rho_{i,j+1} +
   \frac{f^{B\varpi}_{i,j}}{2 \Delta z} \rho_{i,j-1} +
   (f^{Bz}_{i,j} f^{g\varpi}_{i,j} -f^{B\varpi}_{i,j} f^{gz}_{i,j})
   \rho_{i,j}=0,
\end{equation}
where $\Delta z$ and $\Delta\varpi$ are the grid sizes in the $z$
and $\varpi$ axes, respectively, and
\begin{eqnarray}\label{ffs}
  \nonumber f^{B\varpi}_{i,j} \equiv \left(\frac{\partial \Phi_B}{\partial \varpi}\right)_{i,j},\quad
   f^{Bz}_{i,j} \equiv \left(\frac{\partial \Phi_B}{\partial
   z}\right)_{i,j},\\
  \nonumber f^{g\varpi}_{i,j} \equiv \left(\frac{\partial \Phi_g}{\partial \varpi}\right)_{i,j},\quad
   f^{gz}_{i,j} \equiv \left(\frac{\partial \Phi_g}{\partial z}\right)_{i,j},\quad
\end{eqnarray}
must be evaluated (via smoothed finite difference method) with the
new obtained (via SOR method) values of $\Phi_B$ and $\Phi_g$ on the
internal grid points. Numbering the two dimensions of grid points in
a single one-dimensional sequence by $l=(i-2)(N-2)+(j-1)$, equation
(\ref{newrho}) takes the matrix form $\mathbf{A} \mathbf{x}=
\mathbf{b}$ where $\mathbf{A}$ is a sparse matrix similar to the
Fig.~19.0.3 of Press et al.~(2007). We use the bicongugate gradient
method (the routine LINBCG of Press et al.~2007 that is derived from
a set of iterative routines originally written by Anne Greenbaun) to
solve this set of linear algebraic equations with sparse matrix.
Now, knowing the density in all grid points, the current density can
be obtained via
\begin{equation}\label{newj}
   j_{\phi i,j} = \frac{\varpi_i}{f^{B\varpi}_{i,j}}
   \left( \frac{\rho_{i+1,j}- \rho_{i-1,j}}{2\Delta \varpi} +
   f^{g\varpi}_{i,j} \rho_{i,j} \right).
\end{equation}

We use the new obtained values of $\rho$ and $j_{\phi}$ to iterate
the procedure of solving the equation~(\ref{sor}) via SOR, and then
obtaining new values of $\rho$ and $j_{\phi}$ via
equations~(\ref{newrho}) and (\ref{newj}). This iterative method
generates a sequence of improved solutions for $\rho$ and
$j_{\phi}$. The isodensity contours are displayed in the
Fig.~\ref{isoden} for two values of the central density $\rho_c = 2$
and $5$.

\section{Summary and conclusion}

\begin{figure} \epsscale{0.5} \plotone{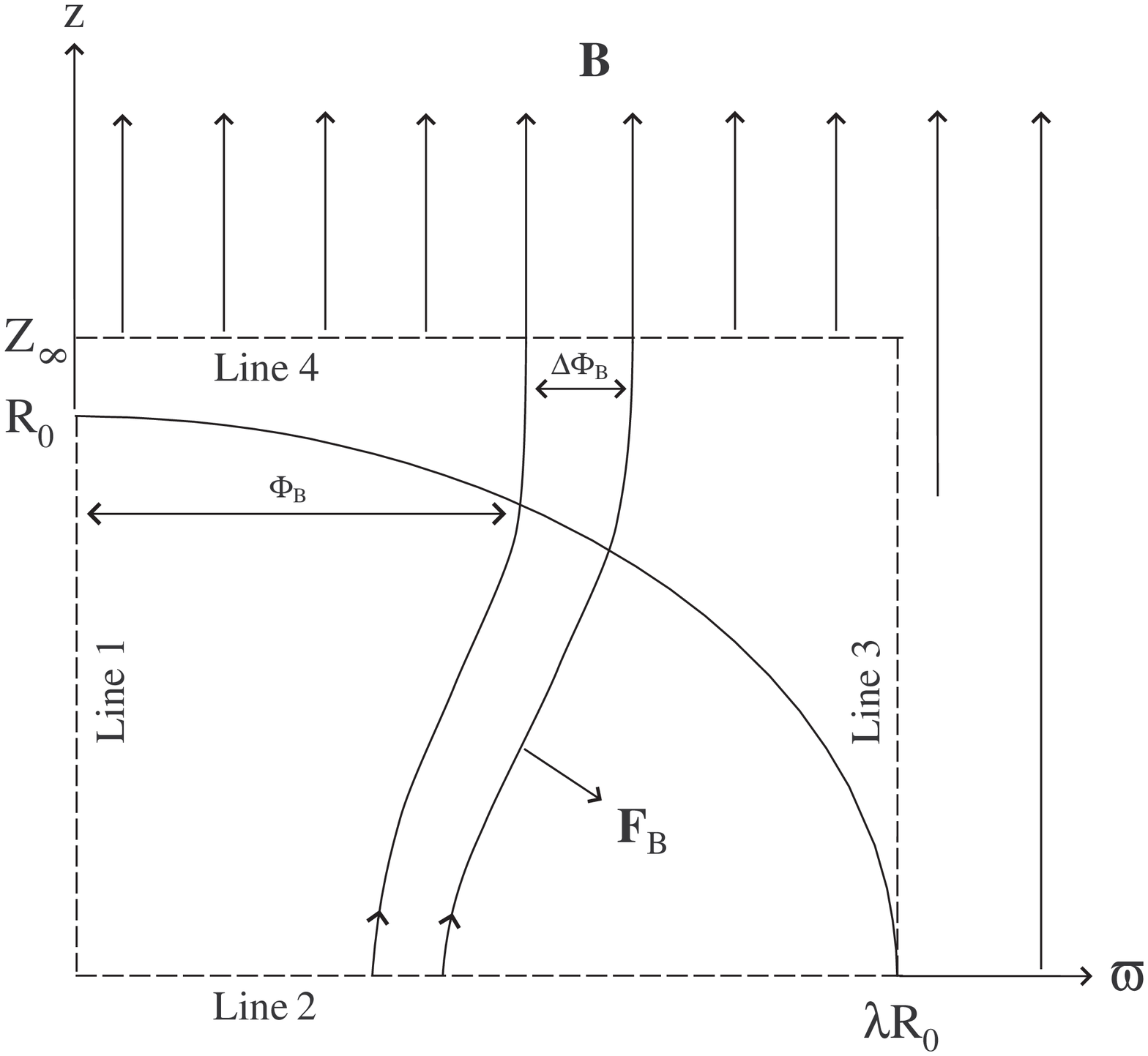}
\caption{One quadrant of a magnetically supported nearly spherical
(oblate with axis ratio $\lambda$) IRDC clump. The field curvature
creates an outward magnetic force, as shown. The boundaries are
depicted by dash lines $1-4$, and the magnetic field in the outer
regions are assumed to be uniform. $\Phi_{B}$ is the magnetic flux
contained within the surface generated by rotating any field line
about the $z$-axis.\label{magnetic}}
\end{figure}

In this work, we investigated the effect of the magnetic field
morphology on the structure of the massive IRDC clumps. To achieve
this goal, we constructed the fundamental equations for an
axisymmetric isothermal oblate IRDC clump, embedded into a constant
external magnetic field, so that the external field is assumed to be
along the $z$ axis of the cylindrical polar coordinates. As shown
schematically in the Fig.~\ref{magnetic}, the radii of the clump in
the direction of the $z$ and $\varpi$ axes are $R_0$ and $\lambda
R_0$, respectively, where $\lambda>1$ is the oblate axis ratio. For
simplicity, we assume that the boundaries at one quadrant of $\varpi
- z$ plane, are specified as four lines~$1-4$, as depicted in the
Fig.~\ref{magnetic}. To solve these equations, we need the boundary
value conditions.

\begin{figure} \epsscale{0.5} \plotone{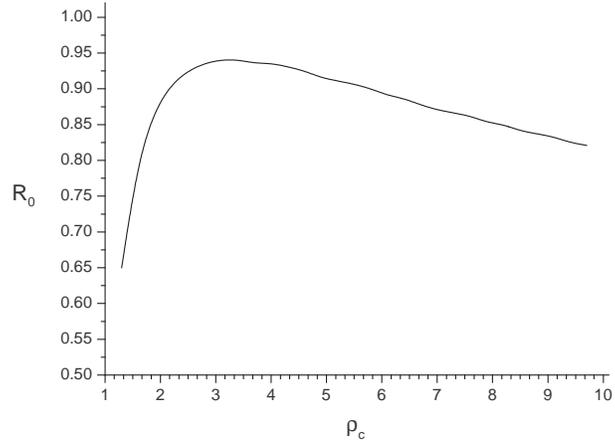}
\caption{Radius of the IRDC clump in the $z$-axis direction for
different values of the central density. The density at $z=R_0$ must
be equal to the inter-clump density
$\rho_{int}=1[\rho]$.\label{rhocR0}}
\end{figure}

We assumed that the density on the boundary line~$1$ is corresponded
with the profile of an infinite layer. In the assumption of infinite
layer, the values of $R_0$ versus different values of the central
density, $\rho_c$, is depicted in the Fig.~\ref{rhocR0}. Similar to
the Bonnor-Ebert spheres in the spherical cases, the density
contrast, $\rho_c/\rho_{int}$, greater than $3.5$ causes to contract
the layer in the $z$-axis direction due to the importance of the
self-gravitational force. To consider the oblateness of the IRDC
clump and departure from infinite layer approximation, we
approximately considered the term $\frac{1}{\varpi}\frac{\partial
}{\partial\varpi}\left(\varpi\frac{\partial
\Phi_{g}}{\partial\varpi}\right)$, and reevaluate the $R_0$, in each
iterative process.

\begin{figure} \epsscale{0.5} \plotone{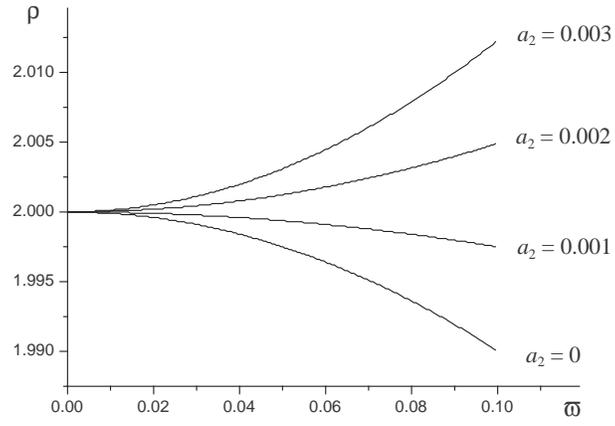}
\caption{Density near the origin of the boundary line~$2$, with
magnetic field morphology $B_{(\varpi,z=0)} =B_{int} \times
\left[(1+\eta)-a_2 \varpi^2\right]$.\label{rho-varpi}}
\end{figure}

In boundary line~$2$, we assume the density and the magnetic field
strength are corresponded with the profiles of an infinite cylinder.
Here, we need a magnetic field morphology. We considered a
polynomial function for the magnetic field morphology. Decreasing of
the magnetic field by moving away from the center implies that the
magnetic force be in the inverse direction of the gravitational
force so that the pressure (and also density) increases as shown in
the Fig.~\ref{rho-varpi}. Here, we chose the magnetic morphology as
given by polynomial equation~(\ref{Bl2}) which appropriately justify
the boundary conditions.

\begin{figure}
\epsscale{.4} \center \plotone{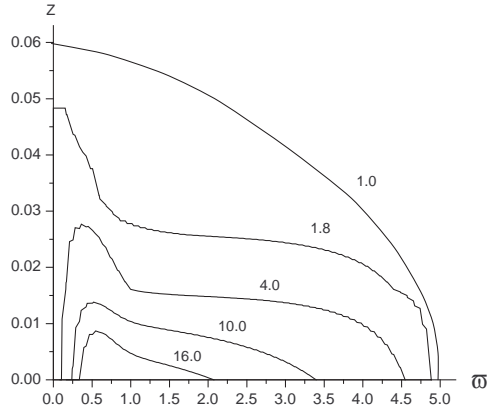}\\{(a)}\\
\epsscale{.4} \center \plotone{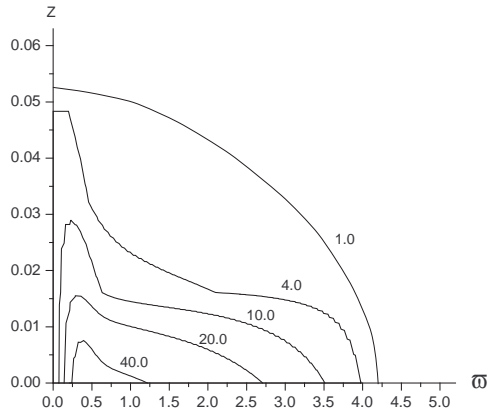}\\{(b)}\\
\caption{Isodensity contours for one quadrant of IRDC clump with (a)
$\rho_c=2$ and (b) $\rho_c=5$.\label{isoden}}
\end{figure}

Boundary lines~$3$ and $4$ are located at the inter-clump medium.
With known boundary value conditions on these four lines, we solved
equations~(\ref{neweq4}) and (\ref{neweq3}), with the
successive-over-relaxation method, to obtain $\Phi_g$ and $\Phi_B$.
Then, we used the obtained values of $\Phi_g$ and $\Phi_B$ to solve
the equations (\ref{neweq1}) and (\ref{neweq2}), with the
bicongugate gradient method, to find new values of $\rho$ and
$j_\phi$. Then, we put these values in equations~(\ref{neweq4}) and
(\ref{neweq3}) and iterate this process. In Fig.~\ref{isoden}, the
isodensity is plotted for two values of the central density $\rho_c
= 2$ and $5$.

In both diagrams of Fig.~\ref{isoden}, we see that, firstly, the
IRDC clump is very elongated along the perpendicular direction of
the external magnetic field lines. Secondly, the choice of the
magnetic field morphology (\ref{Bl2}) leads to the formation of
dense regions around the center. The greater the density of the
central region, the larger the density of these dense regions and
the closer to the $z$ axis. The presence of these dense regions can
lead to the formation of cores at these points. These results are
somewhat consistent with the results of figure~1 of Sanhueza et
al.~(2017), who has found multiple cores near the elongated axis and
\textit{around the center} of the massive clump of IRDC
G028.23-00.19.

We have shown that by choosing the appropriate morphology for the
magnetic field, we can justify the presence of the multiple cores
\textit{around the center} and near the elongated axis of the IRDC
clump G028.23-00.19. In fact, for the general conclusion, not only
the magnitude and direction of the magnetic field within the clumps
of IRDCs must be investigated via suitable methods (e.g., techniques
outlined by Hoq et al.~2017), but also must be searched for in the
multiple cores within them via suitable observational techniques
(e.g., methods outlined by Sanhueza et al.~2017). The oblate
structure of the massive clump of IRDC G028.23-00.19 (with dense
regions \textit{around the center} and along its elongated axis),
obtained from magnetic morphology (\ref{Bl2}), may also be
appropriate for clumps of other IRDCs. Increasing new observational
information from inside of the IRDC clumps (i.e., magnetic field and
positions of dense regions through it), may help us to select other
morphologies for the magnetic field. In any way, the morphology of
the magnetic field affects on the sub-structure of the massive IRDC
clumps. We need more observational information from inside of the
IRDC clumps to completely deduce a suitable concluding remark for
choosing an appropriate morphology for the magnetic field.



\clearpage

\clearpage

\clearpage

\clearpage

\end{document}